\documentclass[twocolumn,showpacs,aps,prd]{revtex4}
\pdfoutput=1
\newcommand{\half}{\frac{1}{2}}

\newcommand{\sxth}{\frac{1}{6}}
\newcommand{\R}{{\bar r}}
\usepackage[dvipsnames]{color}
\newcommand{\revision}[1]{{           {#1}}} %suppress highlight
\begin{document}
%\begin{center}\today\hspace{1 cm}For {\em EPL}\end{center}
\title{Conformal gravity: Newton's constant is not universal}
\author{R. K. Nesbet }
\affiliation{
IBM Almaden Research Center,
650 Harry Road,
San Jose, CA 95120-6099, USA
\begin{center}rkn@earthlink.net\end{center}
}
\date{\today}
%***************** VM screen width ************************************%
\begin{abstract}
Newton's gravitational constant $G$ has been measured to high
accuracy in a number of independent experiments.  For
currently unresolved reasons, indicated values from different
well-designed and thoroughly analyzed experiments differ by more
than the sum of estimated errors.  It has recently been shown
that requiring both Einstein general relativity and the Higgs scalar
field model to satisfy conformal symmetry (local Weyl scaling
covariance) introduces gravitational effects that explain anomalous 
\revision{
galactic rotation, currently accelerating Hubble expansion, 
}
and dark galactic halos, without invoking dark matter.  This implies 
different values $G_n$ and $G_p$ for neutron and proton, respectively, 
but retains the Einstein equivalence principle for test objects 
accelerated by a given gravitational field. Isotopic mass defect $\mu$ 
per nucleon determines independent $G_m$. Thus G differs for each 
nuclear isotope.  Several recent measurements are used here to 
estimate $G_n=6.60216$, $G_p=6.38926$, and  $G_m=-11.60684$ in units 
$10^{-11}m^3kg^{-1}s^{-2}$.
\end{abstract}
%04.20.Cv Fundamental problems and general formalism
%04.80.Cc Experimental tests of gravitational theories
%06.20.Jr Determination of fundamental constants
\maketitle
% body of paper here

\section{Introduction}
\par
Newton's law $F=Gm_am_b/r_{ab}^2$, for the attractive gravitational 
force between two nonoverlapping massive objects at sufficiently large 
center of mass separation $r_{ab}$, is valid for weak interactions 
in general relativity and is the basis for most contemporary models of 
cosmology\cite{WEI72,DOD03}.  Despite a continuing sequence of 
progressively more accurate measurements of Newton's constant $G$,
individual measured values differ by a collective spread ten times
larger than the uncertainty of a typical measurement\cite{SAQ14,SCH14}.
\par A possible resolution of this paradox is considered here. Newton's
law follows from analysis of gravitation in the Schwarzschild metric, 
valid for a spherically symmetric static mass/energy source in general
relativity\cite{WEI72}.  Bare fermion and gauge fields satisfy local
Weyl scaling\cite{WEY18,WEY18a} (conformal) symmetry\cite{DEW64} but 
the Einstein metric tensor and Higgs scalar field do not\cite{MAN06}. 
\revision{  
If conformal symmetry is imposed on 
general relativity\cite{MAK89,MAN90,MAK94,MAN06,MAN12}, 
an external Schwarzschild
} 
solution is retained, but acquires an acceleration term that becomes 
dominant on a galactic scale, implying anomalously large
orbital rotation velocities, as observed.  Conformal gravity preserves 
subgalactic phenomenology while describing observed excessive rotational
velocities in the outer reaches of galaxies, without invoking dark
\revision{ 
matter. The equivalence principle is retained\cite{MAN06,MAN12}, but
$G$ is no longer a universal constant\cite{MAK94,NESM5}.
}
\par Possible formal difficulties with the conformal version of the
Schwarzschild field equation\cite{FLA06,BAV09,YOO13} have been 
discussed and resolved in detail\cite{MAN07,MAN11,NESM5}. 
\revision{  
$G_n$ and $G_p$, for neutron and proton respectively, have distinct 
universal values\cite{MAK94,NESM5}.  Hence any material mass-energy
source produces a gravitational field and effective $G$ that depends on
its neutron/proton ratio.  Taking nuclear binding energy density into  
account, with parameter $G_m$, each nuclear isotope has a unique value 
of $G$.
}
\par Although the equivalence principle is retained for test objects
in a fixed gravitational field\cite{MAN06}, the interaction between
two gravitating masses depends on atomic composition, governed
by neutron/proton ratios and nuclear binding energy.  The  
difference of $G$ for source mass densities with different atomic
composition may account for the continuing failure of accurate
measurements\cite{SAQ14,SCH14} to converge to a unique common value.
\par The present paper summarizes this theoretical argument and applies
it to deduce approximate values of $G_n,G_p,G_m$ from selected 
experimental data.  The inferred values are clearly not equal.
Several current experimental results are found to be consistent. 

\section{Summary of theory}
\par Physical field equations are determined by requiring that
a spacetime action integral, invariant under symmetry operations
such as Lorentz transformations, should be stationary under
infinitesimal variations of all elementary fields, including the metric 
tensor field of Riemann and Einstein\cite{WEI72,NES03,MAN06}.   
Each distinct physical field contributes a Lagrangian density to
the integrand of the action integral.  Variation of the metric tensor
determines  a gravitational field equation with a uniquely defined 
metric functional derivative tensor for each field. 
\par Multiplying all distinct field amplitudes by specified powers of a
real differentiable function $\Omega(x)$ defines local Weyl (conformal) 
scaling\cite{WEY18,WEY18a}.  Invariance of the action integral under 
local Weyl scaling defines conformal symmetry.
Fermion and gauge boson fields satisfy conformal symmetry\cite{DEW64},
but Einstein gravity and the standard Higgs model do not. Postulated 
universal conformal symmetry\cite{NES13} removes implied inconsistency  
by modifying the Lagrangian densities of both the metric
tensor\cite{WEY18,MAN06} and the Higgs scalar field\cite{NESM1,NESM5}. 
\par Conformal Higgs theory\cite{NESM1,NESM2} is a uniquely 
defined version of standard scalar field theory\cite{MAN06}.
Mass and dark energy must be dynamical quantities resulting from field
interactions\cite{NES13}.  Conformal symmetry causes the Weyl 
gravitational field to drop out of the theory of cosmic evolution, but 
introduces a gravitational term for the Higgs field which has been 
shown to account for observed Hubble expansion, including an 
explanation of dark energy\cite{NESM1,NESM2,NES13}.  The universal 
conformal symmetry postulate has been shown to explain excessive 
galactic rotation velocities\cite{MAN97,MAO11,MAO12,OAM12}, Hubble 
expansion with positive acceleration\cite{NESM1,NESM2}, and dark 
galactic halos\cite{NESM3}, without evoking dark matter, while 
deriving dark energy\cite{NESM5}.

\par The conformal Lagrangian density combines two gravitational terms:
conformal gravity (CG) ${\cal L}_g$\cite{WEY18} and the conformal 
Higgs model (CHM) ${\cal L}_\Phi$\cite{NESM5}. 
The field equation adds the two metric functional derivatives
$X_g^{\mu\nu}+X_\Phi^{\mu\nu}=\half\Theta_m^{\mu\nu}$,
for source energy-momentum $\Theta_m^{\mu\nu}$.
Defining mean source density ${\bar d}$ and
residual density ${\hat d}=d-{\bar d}$, and assuming
$\Theta_m^{\mu\nu}(d)\simeq
 \Theta_m^{\mu\nu}({\bar d})+\Theta_m^{\mu\nu}({\hat d})$,
decoupled solutions for $r\leq \R$ of the two equations
\begin{eqnarray} \label{Twoeqs}
X_\Phi^{\mu\nu}=\half\Theta_m^{\mu\nu}({\bar d}),
X_g^{\mu\nu}=\half\Theta_m^{\mu\nu}({\hat d})
\end{eqnarray}
imply a solution of the full equation\cite{NESM5}. This requires
a composite hybrid metric\cite{NESM5}  such as
\begin{eqnarray} \label{xmet}
ds^2=-B(r)dt^2+a^2(t)(\frac{dr^2}{B(r)}+r^2d\omega^2), 
\end{eqnarray}
combining Schwarzschild potential $B(r)$ and Friedmann scale factor 
$a(t)$. Solutions are made consistent by fitting parameters to boundary 
conditions\cite{NESM5} and setting cosmic curvature constant $k=0$, 
justified by currently observed data. 
\par The classical solution $B(r)=1-2\beta/r$ determines $\beta=Gm/c^2$,
for $m=\int_0^\R q^2d(q)dq$, which defines gravitational constant $G$.
The conformal $q^4$ integral for $\beta$\cite{MAK94} depends on 
residual mass density\cite{NESM5}, whose $q^2$ integral vanishes by 
definition, consistent with vanishing of the Weyl tensor for a uniform 
source.  It cannot be interpreted to define a universal constant $G$. 

\section{Static, spherically symmetric source density}
\par In the relativistic Schwarzschild model, valid for an isolated 
spherical source density confined inside radius $\R$,
the standard gravitational potential in the source-free exterior 
space $r\geq{\R}$ is $B(r)=1-2\beta/r$. 
For integrated source mass $m$, $\beta=Gm/c^2$ where $G$ is Newton's 
constant.  Mannheim and Kazanas\cite{MAK89,MAK94} showed that conformal 
gravity replaces this by $B(r)=\alpha-2\beta/r+\gamma r-\kappa r^2$.  
A solution of the full tensorial field equation requires 
$\alpha^2=1-6\beta\gamma$\cite{MAK91,BAV09}.
\revision{
The $q^2$ integral for parameter $\beta$ is replaced by
$2\beta=\sxth\int_0^\R q^4{\hat d}(q)dq$\cite{MAK89,MAK94,MAN06}, no
longer strictly proportional to integrated mass $m$. $G=\beta c^2/m$ 
depends on residual mass density\cite{NESM5}.  It may be reduced
by orders of magnitude from $G$ calculated from the full mass density.
}
\par For orbital velocity $v$, such that $v^2=\half r \frac{dB}{dr}$, 
a circular orbit of a test particle is stabilized by centripetal radial
acceleration $v^2/r$.  The implied conformal extension of Kepler's law,
\begin{eqnarray}
v^2=\beta/r+\half\gamma r-\kappa r^2,
\end{eqnarray}
has been fitted to orbital velocity data for 138 
galaxies{\cite{MAN97,MAO11,MAO12,OAM12}.  This determines 
parameters $\beta=N^*\beta^*,\gamma=\gamma_0+N^*\gamma^*, \kappa$, for
each galactic mass $M=N^*M_\odot$ in solar units, defining four 
parameters $\beta^*,\gamma^*,\gamma_0,\kappa$.
More recently, galactic rotation data has been found to define
a function of classical Newtonian acceleration\cite{MLS16} that 
requires $\gamma$ to be independent of galactic mass\cite{NESM7},
as predicted by the conformal Higgs model\cite{NESM2,NESM5}. 

\section{Conformal gravitational constant}
\par In standard theory, the differential equation for $B(r)$ is of 
second order, so that the integral solution for $\beta$ of the form
$\int q^2 d(q)dq$ is strictly proportional to the integrated mass. 
\revision{
However, conformal theory replaces this by an integral of the form  
$\int q^4 {\hat d}dq$, for a fourth order differential 
equation\cite{MAK94}. This depends on residual 
density ${\hat d}$\cite{NESM5}, which integrates to zero over a closed 
sphere if weighted by $q^2$. 
Residual density is determined by mass-energy density 
but the $q^4$ integral is not proportional to integrated mass.
}   
The residual mass integral may be reduced by orders of magnitude
from $\beta$ for the full source mass density, 
accounting for the long-questioned small value of implied $G$.  
However, the equivalence principle, 
due to full mass following a geometrical geodesic, is maintained.
\par As argued by Mannheim\cite{MAN06,MAN12}, the requirement for
Newton's formalism to be valid on a macroscopic scale is that   
each atom should produce a gravitational potential proportional to
$1/r$, since only the sum of such terms at very large distance
(in atomic units) is measured.  In fact, the true basic scale is
nanoscopic, since all stable matter observed in natural conditions is
composed of neutrons and protons.  Assuming global charge neutrality,
electron mass should be added to each proton mass, and nuclear binding
energy should also be included.  Chemical energy may be relevant for
absolute precision.
\par Analysis at the nanoscopic level\cite{NESM5} does not imply 
identical neutron and proton internal source densities
$d_n(r)$ and $d_p(r)$.  There does not appear yet to be an accurate 
prediction of these densities from QCD.  Proton and neutron masses are 
\revision{
different, and there is no reason to assume the $\beta$ integral 
$\int q^4 {\hat d}(q)dq$ to be explicitly proportional to the mass 
integral\cite{MAK94}.  Hence definition $G=\beta c^2/m$ can be 
expected to produce two different fundamental constants $G_n$ and 
$G_p$, for neutron and proton respectively\cite{MAK94,NESM5}.
}
\section{Recent accurate measurements}
Conformal gravity retains the mathematical formalism of standard
general relativity.  A test particle moving in a fixed gravitational 
field follows a geodesic of that field, which ensures the Einstein 
equivalence principle\cite{MAN06,MAN12}. More generally, an experiment 
that measures acceleration of a negligibly small mass $m_a$ in the 
far gravitational field of a much larger mass $m_b$ is a specific 
measurement of $G_b$ only. This is appropriate to a recent 
atomic-physics experiment in which two isotopes of strontium
are accelerated by the earth's gravitational field.  The equivalence 
principle is verified to one part in $10^7$\cite{TMP14}. 
This Sr isotope experiment has been reconfigured as a measurement of 
$G$ using rubidium atoms accelerated by a source mass of heavy 
tungsten alloy cylinders\cite{RSC14}.  Error analysis gives 
$G=6.67191(99)$ in units $10^{-11}m^3kg^{-1}s^{-2}$.
\par Another recent experiment uses laser interferometry to measure
deflection of two free-hanging pendulum masses by a much larger
tungsten alloy source mass\cite{PAF10}.  The gravitational constant
is found to be $G=6.67234(14)$.
These two experiments, with overlapping error bars, are characterized
by a massive field source, with negligible deflected test mass.  
Their results will be considered here as measurements of $G$ for the
source tungsten alloy.  For element W with the natural isotopic
abundance, the neutron and proton fractions are $f_n=0.598, f_p=0.402$. 
\par A somewhat earlier torsion-balance experiment was designed to
eliminate or greatly reduce uncertainty due to the suspending
torsion fiber, to the mass and geometry of the pendulum probe, and
to background noise\cite{GAM00}.  The 1.5mm thick pyrex pendulum
mass is negligible compared with field source mass due to
four 8.14kg stainless steel alloy (SS316) spheres.  Reported
$G=6.674215(092)$ is considered here to
measure $G$ for the SS316 alloy.  Averaged over alloy composition,
$f_n=0.535, f_p=0.465$.
\par An experiment using the same material for source and test masses
measures $G$ for that material.
Such an experiment was reported in 2001\cite{QSR01}, using
a torsion-strip balance in two different measurement modes.
Both source and test masses were made from the same Cu alloy.
The apparatus was rebuilt and error analysis refined to give a
more recent result\cite{QPS13,QSPD14}, $G=6.67554(16)$.
Averaged $f_n=0.544, f_p=0.456$.
\par Another experiment, using a compensated torsion balance designed
to reduce or eliminate systematic error, used a suspended mass probe
of 500g Cu, with two large cylindrical source masses, approximately 
27kg each\cite{AAF03}.  Results were obtained for two different
source materials.  Measured $G(SS316)=6.67392(49)$\cite{AAF03} is 
consistent with $G=6.674215(092)$\cite{GAM00}. 
Measured $G(Cu)=6.67385(26)$\cite{AAF03} is inconsistent with
$G=6.67554(16)$\cite{QSPD14}.  Measurements of $G$ cited here are shown 
in Figure I, including the 2010 CODATA value 6.67384(80)\cite{COD10}.
%***************** VM screen width ************************************%
\par Evaluation of G requires computing Schwarzschild parameter
$\beta$ using accurate intra-nuclide energy density\cite{NESM5}.
Then $G=\beta c^2/M$.
Empirical data used here is neutron mass $m_n$, proton plus electron 
mass $m_p$, and mass defect $\Delta$ per nucleon $\mu=\Delta/Ac^2$.
Constants are $N_p=Z, N_n=N, A=N_p+N_n$ and $M=N_n m_n+N_p m_p-A\mu$.
\revision{
$M$ and $N_n$ here are summed over stable isotopes for each element, 
using known abundance fractions.  Neutron and proton fractions are
computed for weighted $N_n$.  Assuming additive contributions to  
$\beta=MG/c^2$, 
\begin{eqnarray} \label{beta}
\beta c^2/A=f_nm_nG_n+f_pm_pG_p-\mu G_m=MG/A
\end{eqnarray}  
for each element.  
Using data for Fe\cite{GAM00}, Cu\cite{QSPD14}, W\cite{PAF10} 
in units $10^{-11}m^3kg^{-1}s^{-2}$, Eq.(\ref{beta}) implies
}
\begin{eqnarray}
Fe:
0.53960G_n+0.46867G_p-0.009434G_m=6.66648\nonumber\\                             
Cu:
0.54886G_n+0.45942G_p-0.009397G_m=6.66810\nonumber\\             
W:
0.60277G_n+0.40556G_p-0.008595G_m=6.67055.   
\end{eqnarray} 
Hence $G_n=6.60216, G_p=6.38926, G_m=-11.60684$.\\
Since $G_m$ depends on residual energy density, which integrates 
to zero, its sign is not predetermined.

\setlength{\unitlength}{0.015cm}
\begin{center}
\begin{picture}(600,800)(0,0)\thicklines
\put(000,000){\line(1,0){600}}
\put(000,800){\line(1,0){600}}

\put(000,000){\line(0,1){800}}
\put(600,000){\line(0,1){800}}
 \put(-20,-30){6.670}
 \put(000,800){\line(0,-1){10}}
 \put(080,-30){6.671}
 \put(100,000){\line(0,1){10}}
 \put(100,800){\line(0,-1){10}}
 \put(180,-30){6.672}
 \put(200,000){\line(0,1){10}}
 \put(200,800){\line(0,-1){10}}
 \put(280,-30){6.673}
 \put(300,000){\line(0,1){10}}
 \put(300,800){\line(0,-1){10}}
 \put(380,-30){6.674}
 \put(400,000){\line(0,1){10}}
 \put(400,800){\line(0,-1){10}}
 \put(480,-30){6.675}
 \put(500,000){\line(0,1){10}}
 \put(500,800){\line(0,-1){10}}
 \put(580,-30){6.676}
 \put(600,000){\line(0,1){10}}
%%%%%## 
 \put(092,700){\line(1,0){198}} 
 \put(092,695){\line(0,1){10}}
 \put(191,690){\line(0,1){20}}
 \put(290,695){\line(0,1){10}}
 \put(010,700){\cite{RSC14}W}
%%%%## 
 \put(220,600){\line(1,0){28}} 
 \put(220,595){\line(0,1){10}}
 \put(234,590){\line(0,1){20}}
 \put(248,595){\line(0,1){10}}
 \put(010,600){\cite{PAF10}W}
%%%%## 
 \put(412.3,500){\line(1,0){18.4}} 
 \put(412.3,495){\line(0,1){10}}
 \put(421.5,490){\line(0,1){20}}
 \put(430.7,495){\line(0,1){10}}
 \put(010,500){\cite{GAM00}SS}
%%%%## 
 \put(538,400){\line(1,0){36}} 
 \put(538,395){\line(0,1){10}}
 \put(554,390){\line(0,1){20}}
 \put(570,395){\line(0,1){10}}
 \put(010,400){\cite{QSPD14}Cu}
%%%%## 
 \put(359,300){\line(1,0){52}} 
 \put(359,295){\line(0,1){10}}
 \put(385,290){\line(0,1){20}}
 \put(411,295){\line(0,1){10}}
 \put(010,300){\cite{AAF03}Cu}
%%%%## 
 \put(343,200){\line(1,0){98}} 
 \put(343,195){\line(0,1){10}}
 \put(392,190){\line(0,1){20}}
 \put(441,195){\line(0,1){10}}
 \put(010,200){\cite{AAF03}SS}
%%%%## 
 \put(304,100){\line(1,0){160}} 
 \put(304,095){\line(0,1){10}}
 \put(384,090){\line(0,1){20}}
 \put(464,095){\line(0,1){10}}
 \put(010,100){\cite{COD10}}
%%%%## 
\end{picture}
\end{center}
%%%%##
\vspace{1 mm}
\begin{center}
Figure I.
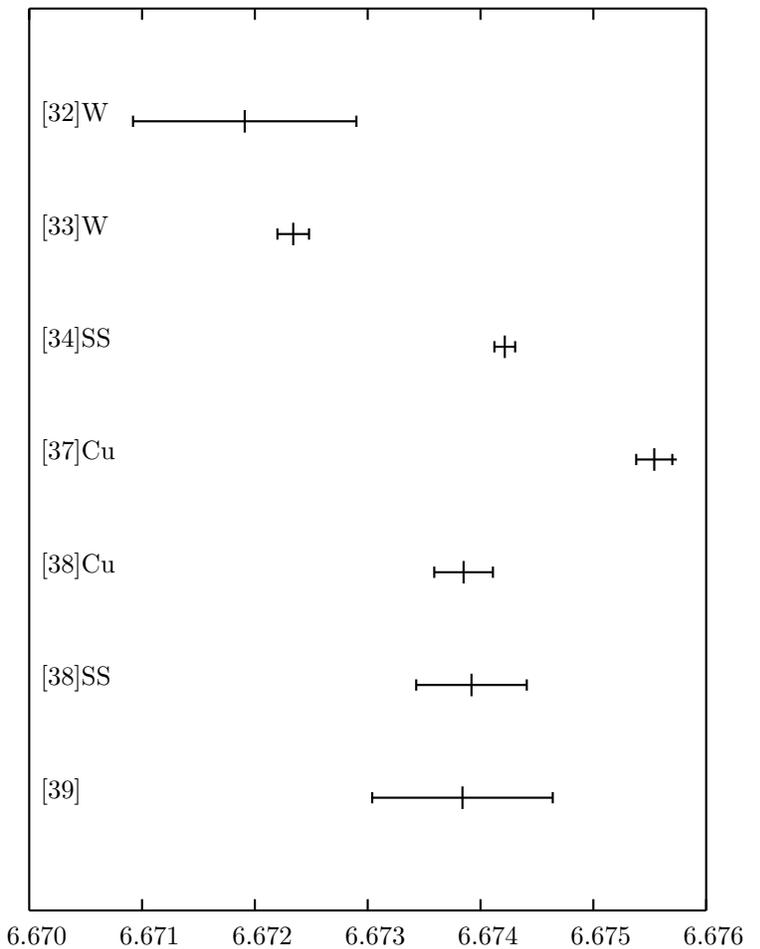
Newton G in units $10^{-11}m^3kg^{-1}s^{-2}$ 
\end{center}

\section{Conclusions}
\par Conformal theory indicates that Newton's constant $G$ must depend
on the atomic composition of any massive gravitational field source. 
Neutron and proton masses and nuclear binding energy contribute 
separately. This appears to resolve some but not all of the apparent
inconsistencies among recent experimental measurements of assumed
universal constant $G$.  The situation might be clarified by measuring 
$G$ in the same apparatus for different source mass materials.   
Definitive results are more likely to be achieved when the test mass
is negligible compared with the gravitational field source.
%***************** VM screen width ************************************%


\begin{thebibliography}{99}
\bibitem{WEI72} S. Weinberg,
{\it Gravitation and Cosmology:
Principles and Applications of the General Theory of Relativity}
(Wiley, New York, 1972).
\bibitem{DOD03} S. Dodelson,
{\it Modern Cosmology}
(Academic Press, New York, 2003).
\bibitem{SAQ14} C. Speake and T. Quinn,
{\it Physics Today} {\bf 67}, 27 (2014).
\bibitem{SCH14} S. Schlamminger,
{\it Nature} {\bf 510}, 478 (2014).
\bibitem{WEY18} H. Weyl,
{\it Sitzungber.Preuss.Akad.Wiss.} {26}, 465 (1918).
\bibitem{WEY18a}
{\it Math.Zeit.} {\bf 2}, 384 (1918).
\bibitem{DEW64} B. S. DeWitt, in
{\it Relativity, Groups, and Topology},
C. DeWitt and B. S. DeWitt, eds.
(Gordon and Breach, New York, 1964).
\bibitem{MAN06} P. D. Mannheim,
{\it Prog.Part.Nucl.Phys.} {\bf 56}, 340 (2006).
%(arXiv:astro-ph/0505266v2)
\bibitem{MAK89} P. D. Mannheim and D. Kazanas,
{\it ApJ} {\bf 342}, 635 (1989).
\revision{
\bibitem{MAK94} P. D. Mannheim and D. Kazanas,
{\it Gen.Rel.Grav.} {\bf 26}, 337 (1994).
}
\bibitem{MAN90} P. D. Mannheim,
{\it Gen.Rel.Grav.} {\bf 22}, 289 (1990).
\bibitem{MAN12} P. D. Mannheim,
{\it Found.Phys.} {\bf 42}, 388 (2012).
\bibitem{NESM5} R. K. Nesbet,
{\it Europhys.Lett.} {\bf 131}, 10002 (2020).
\bibitem{FLA06} E. E. Flanagan,
{\it Phys.Rev.D} {\bf 74}, 023002 (2006).
\bibitem{BAV09} Y. Brihaye and Y. Verbin,
{\it Phys.Rev.D} {\bf 80}, 124048 (2009).
\bibitem{YOO13} Y. Yoon,
{\it Phys.Rev.D} {\bf 88}, 027504 (2013).
\bibitem{MAN07} P. D. Mannheim,
{\it Phys.Rev.D} {\bf 75}, 124006 (2007).
\bibitem{MAN11} P. D. Mannheim,
{\it Gen.Rel.Grav.} {\it 43}, 73 (2011).
\bibitem{NES03} R. K. Nesbet,
{\it Variational Principles and Methods in Theoretical Physics
 and Chemistry}
(Cambridge Univ. Press, New York, 2003).
\bibitem{NES13} R. K. Nesbet,
{\it Entropy} {\bf 15}, 162 (2013).
\bibitem{NESM1} R. K. Nesbet,
{\it Mod.Phys.Lett.A} {\bf 26}, 893 (2011),
%(arXiv:0912.0935v3 [physics.gen-ph]).
\bibitem{NESM2} R. K. Nesbet,
{\it Europhys.Lett.} {\bf 125}, 19001 (2019).
\bibitem{MAN97} P. D. Mannheim,
{\it ApJ} {\bf 479}, 659 (1997). 
\bibitem{MAO11} P. D. Mannheim and J. G. O'Brien,
{\it Phys.Rev.Lett.} {\bf 106}, 121101 (2011).
\bibitem{MAO12} P. D. Mannheim and J. G. O'Brien,
{\it Phys.Rev.D} {\bf 85}, 124020 (2012).
\bibitem{OAM12} J. G. O'Brien and P. D. Mannheim,
{\it MNRAS} {\bf 421}, 1273 (2012).
\bibitem{NESM3} R. K. Nesbet,
{\it Europhys.Lett.} {\bf 109}, 59001 (2015).
\bibitem{MAK91} P. D. Mannheim, and D. Kazanas,
{\it Phys.Rev.D} {\bf 44}, 417 (1991).
\bibitem{MLS16} S. S. McGaugh,  F. Lelli and J. M. Schombert,
{\it Phys.Rev.Lett.} {\bf 117}, 201101 (2016).
\bibitem{NESM7} R. K. Nesbet,
{\it MNRAS} {\bf 476}, L69 (2018).
\bibitem{TMP14} M. G. Tarallo et al,
{\it Phys.Rev.Lett.} {\bf 113}, 23005 (2014).
\bibitem{RSC14} G. Rosi et al,
{\it Nature} {\bf 510}, 518 (2014).
\bibitem{PAF10} H. V. Parks and J. E. Faller,
{\it Phys.Rev.Lett.} {\bf 105}, 110801 (2010).
\bibitem{GAM00} J. H. Gundlach and S. M. Merkowitz,
{\it Phys.Rev.Lett.} {\bf 85}, 2869 (2000).
\bibitem{QSR01} T. J. Quinn et al,
{\it Phys.Rev.Lett.} {\bf 87}, 111101 (2001).
\bibitem{QPS13} T. J. Quinn et al,
{\it Phys.Rev.Lett.} {\bf 111}, 101102 (2013).
\bibitem {QSPD14} T.J. Quinn et al,
{\it Phil.Trans.R.Soc.A} {\bf 372}, 32 (2014).
\bibitem{AAF03} T. R. Armstrong and M. P. Fitzgerald,
{\it Phys.Rev.Lett.} {\bf 91}, 201101 (2003). 
\bibitem{COD10} P. J. Mohr et al,
{\it Rev.Mod.Phys.} {\bf 84}, 1527 (2010).
\end{thebibliography}
\end{document}